\documentclass[12pt,thmsa]{article}
\usepackage{sw20lart}

\input{tcilatex}
\begin{document}

\author{R. Holman and T.W. Kephart \and Department of Physics \& Astronomy, \and %
Vanderbilt University, Nashville, TN 37235 USA}
\title{${}SO(8)$ and $G_{2}\times G_{2}$ $N=1$ Supergravities in $D=6$ }
\date{15 May 1998}
\maketitle

\begin{abstract}
We construct the $N=1$ supergravity analog of the Green-Schwarz and
heterotic superstring theories in $10D$. We find the $SO(8)$ theory
previously found by compactification of $10D$ on $K3$. We also find eleven $%
G_{2}\times G_{2}$ theories, two with symmetric matter content. It is not
obvious how tenof the $G_{2}\times G_{2}$ theories can be gotten from $10D$.
\end{abstract}

Ten-dimensional superstrings \cite{jhs}, \cite{GS}, \cite{heterotic}, \cite
{W+A-G} are one of the most important discoveries of our time. They have led
to fundamental advance in the way we perceive the relationships between
particle physics, condensed matter physics, gravity and mathematics. 

The zero slope limit of the Green-Schwarz $S0(32)$ open string theory in 10D
is an $N=1$ supergravity theory, and is interesting in its own right. The
important observation \cite{GS} that the leading gauge anomaly\cite{FK and
TS} in $SO(N)$ is proportional to $(N-32)$ means only $SO(32)$ is possible.
The development of the Green-Schwarz mechanism restored consistency to the
theory, but only for $S0(32)$ where in addition $496$ gauge group generators
(vector fields) are required to cancel the perturbative gravitational
anomaly. Once the $SO(32)$ open string theory and then the heterotic theory
were established, much progress was made by various methods of
compactification, and by investigating generalized solitons and instantons
(p-branes, D-branes, etc.) and other nonperturbative aspects including
dualities in these and related theories. By extending the theories to $11$
and $12$ dimensions in the so-called $M$ and $F$ theories relations were
discovered among the $10D$ theories.

Recently $D=6,N=1$ theories have taken center stage \cite{6Drefs}since much
can be learned from compactifying $10D$ theories to $6D$ on the $K3$
manifold or related orbifolds. Several new classes of theories with extended
guage groups \cite{witten6d}\cite{schwarz6d}\cite{erler}have emerged in $6D.$

Here our aims are rather modest in that we wish to examine the closest $6D$ $%
N=1$ supergravity analog of the $10D$ open $0(32)$ and heterotic
superstrings. First we observe that the leading one-loop gauge anomaly in $%
D=6$ is proportional to $N-8$ for $SO(N)$ gauge groups. (Likewise one can
show that the leading 2-loop\cite{FKY} and 3-loop \cite{CK}perturbative
gauge anomalies also have factors of $N-8$.) The $SO(8)$ that cancels this
anomaly has $28$ generators, and one might expect this fact to play a
prominent role in the cancellation of both the leading perturbative
gravitational anomaly and the global gravitational anomaly. However, the
situation here is apparently more subtle than $D=10,N=1$ case, since the
selection of the allowed irreducible representations in $6D$ supergravity is
more rich and including hypermultiplets and tensor multiplets. Before we
move on to trying to construct the $SO(8)$ supergravity, we observe that
there is another rank for gauge group with 28 generators, namely $%
G_{2}\times G_{2}$, and that both $SO(8)$ and $G_{2}\times G_{2}$ have
maximal subgroups $O(4)\times O(4)$. This is rather reminiscent of the $%
O(16)\times O(16)$ maximal subgroups of $O(32)$ and $E_{8}\times E_{8}$. 
\footnote{%
Also it may be worthy to note the curiosity that the number of generators in
the $10D$ theory, i. e., $496$, is a perfect number with deep
number-theoretic significance. The sequence of perfect numbers is $%
1,6,28,496 $, \ldots While the significance of $28$ generators in $6D$ will
remain obscure here, it is interesting to also guess that in $4D$ there may
be an $S0(4)$ or equivalently $SU(2)\times SU(2)$ theory with $6$
generators, and in $2D$ a related $SO(2)$ theory.}

Let us now proceed with the construction of the anomaly-free $SO(8)$ $N=1$
supergravity in $6D$. Perturbative anomaly cancelation in $6D$ is understood
by examining the box diagram with an even numbers of guage field and
graviton legs and massless fermions in the loop. These anomalies are either
pure guage, pure gravitational or mixed. Anomaly freedom requires the
leading pure anomalies to either vanish or cancel and the remaining
anomalies to cancel by the Green-Schwarz mechanism.

The most convenient way to classify the massless particle representations in 
$N=1$ $6D$ supergravity is via their $O(4)$ $\sim $ $SU(2)\times SU(2)$
little group \cite{schwarz6d} where the spectrum contains:

$(3,3)+2(2,3)+(1,3)$ graviton

$(3,1)+2(2,1)+(1,1)$ tensor multiplet

$(2,2)+2(1,2)$ vector multiplet

$2(2,1)+4(1,1)$ hypermultiplet

As usual there is a single graviton supermultiplet but we can let $n_{T}$, $%
n_{V}$, and $n_{H}$ be the numbers of tensor, vector, and hypermultiplets.
We restrict ourselves to gauge group $G$ either $SO(8)$ or $G_{2}\times
G_{2} $and so $n_{V}=dimG=28$. The hypermultiplets are in some
representation $R$ of $G$, so $n_{H}=dimR$. We take $n_{T}=1$, as is
expected from $10D$ superstring compactification, but note that this
restriction could be relaxed \cite{tensor}.

The total perturbative anomaly polynomial $I_{6D}$ \cite{W+A-G}\cite{GSW}%
\cite{BKSS} has contributions from gravitons, dilatons, guaginos, and matter
fermions. This result has been expressed in a form that meets our needs in 
\cite{erler}:

\begin{center}
$
\begin{array}{lll}
24i(2\pi )^{3}I_{6D} & =\frac{1}{240}(272-n_{H})trR^{4}-\frac{1}{192}%
(16+n_{H}) &  \\ 
& -\frac{1}{4}trR^{2}\left[ \sum_{G}TrF_{G}^{2}-\
\sum_{i,G}n_{H}^{i}tr_{R^{_{i}}}trF_{G}^{2}\right] &  \\ 
& +\left[
\sum_{G}TrF_{G}^{4}-\sum_{i,G}n_{H}^{i}tr_{R^{_{i}}}trF_{G}^{4}\right] &  \\ 
& -6\sum_{i,i^{\prime };G,G^{\prime }}n_{H}^{i,i^{\prime }}\left(
tr_{R^{_{i}}}trF_{G}^{2}\right) \left( tr_{R^{_{i^{\prime }}}}trF_{G^{\prime
}}^{2}\right) & 
\end{array}
$
\end{center}

Here $n_{H}^{i}$ is the number of hypermultiplets in the representation R$%
_{i}$ of the guage group factor G. The last term in this equation is absent
when $G=O(8)$, but for $G$=$G_{2}\times G_{2}$ it survives with $%
n_{H}^{i,i^{\prime }}$ the number of hypermultiplets in the representation $%
(R_{i},R_{i^{\prime }})$ or $G_{2}\times G_{2}.$ The Tr traces are in the
adjoint representation and the $tr_{R^{_{i}}}$ traces are taken in the $R_{i}
$ representation. For any guage group we can write all traces in terms the
unindexed $tr$ trace for in the fundamental representation. For the case at
hand we need the following relations:

$
\begin{array}{ll}
TrF_{SO(8)}^{2} & =6trF_{SO(8)}^{2} \\ 
TrF_{SO(8)}^{4} & =3\left( trF_{SO(8)}^{2}\right) ^{2} \\ 
TrF_{G_{2}}^{2} & =4trF_{G_{2}}^{2} \\ 
TrF_{G_{2}}^{4} & =\frac{5}{2}\left( trF_{G_{2}}^{2}\right) ^{2} \\ 
trF_{G_{2}}^{2} & =\frac{1}{4}\left( trF_{G_{2}}^{2}\right) ^{2}
\end{array}
$

and by triality,

$trF_{SO(8)}^{2}=trF_{8_{v}}^{2}=trF_{8_{s}}^{2}=trF_{8_{c}}^{2},$

and

$trF_{SO(8)}^{4}=trF_{8_{v}}^{4}=trF_{8_{s}}^{4}=trF_{8_{c}}^{4}.$

The $I_{6D}$ anomaly can be made to vanish by requiring the following three
conditions be satisfied:

\begin{enumerate}
\item  The leading gravitational anomaly must cancel directly, hence we
demand $n_{H}=272$.

\item  The leading guage anomaly must also vanish or cancel directly. This
happens trivially for $G_{2}\times G_{2}$ since $G_{2}$ has no forth order
Casimir invariant.

For $SO(8)$ the leading anomaly is zero for the adjoint representation, so
there is no guagino contribution but the vector and spinor irreps all have
nonvanishing forth order Casimir and so the only way for the leading guage
anomaly to vanish in an $SO(8)$ theory with lowest level matter
hypermultiplets is to allow only guage group singlet hypermultiplets. Hence
we have a consistent $SO(8)$ theory with $272$ singlet matter
hypermultiplets. This result agrees with the limiting case of enhanced
symmetry theories with guage group $SO(8+n)\times Sp(n)$ with $24-n$ units
of instanton number embedded in $SO(32)$ \cite{small}. These theories have
hypermultiplets in the $\frac{1}{2}(8+n,2n)+\frac{24-n}{2}\left( 1,2n\right) 
$ representation of $SO(8+n)\times Sp(n)$ plus the antisymmetric
representation of $Sp(n).$ The total number of singlet hypermultiplets is $%
20+\frac{1}{2}\left( n-24\right) \left( n-21\right) $ plus one associated
with the antisymmetric tensor field of $Sp(n).$ If we let $n=0$ the guage
group becomes $SO(8)$ and the number of singlet hypermultiplets is precisely 
$272.$ This also agrees with the limiting case of the infinite sequence of
models found by Schwarz \cite{schwarz6d}where $G=SO(8+2n)\times Sp(n)$ and
the $\ $hypermultiplets are in the representation $(2n+8,2n)+272(1,1)$. Here

again for $n=0$ we have $G=SO(8)$ with $272$ singlet hypermultiplets.

\item  The remaining nonleading and mixed anomalies must factorize into a
form proportional to

$\left[ trR^{2}-\sum_{G}\alpha _{G}^{(1)}trF_{G}^{2}\right] \left[
trR^{2}-\sum_{G^{\prime }}\alpha _{G^{\prime }}^{(2)}trF_{G^{\prime
}}^{2}\right] $ in order that they can be canceled by the Green-Schwarz
mechanism.

For $SO(8)$ this adds no new constraint since we can only have guage singlet
hypermultiplets. For $G_{2}\times G_{2}^{\prime }$ this constraint is
nontrivial (we add a prime to distinguish the two $G_{2}$s). $\ $If we have $%
n_{7}$ hypermultiplets in the (7,1) representation and $n_{7}^{\prime }$
hypermultiplets in the (1,7) representation of $G_{2}\times G_{2}^{\prime }$
then \cite{erler}

$\alpha _{G}^{(1)}=\alpha _{G^{\prime }}^{(1)}=1,$

$\alpha _{G}^{(2)}=\frac{n_{7}-10}{6},$

and

$\alpha _{G^{\prime }}^{(2)}=\frac{n_{7}^{\prime }-10}{6}.$

If we have no further nontrivial hypermultiplet representations then the
factorization constraint implies that the $\alpha $s satisfy the following
relation:

$\alpha _{G}^{(1)\ }\alpha _{G^{\prime }}^{(2)}+\alpha _{G}^{(2)}\alpha
_{G^{\prime }}^{(1)}=0.$

This leads to

$n_{7}+n_{7}^{\prime }=20$

and so the number of singlets is N$_{s}=272-7\times 20=132.$ It is
interesting to note that we can have either a symmetric matter content
between $G_{2}$ and $G_{2}^{\prime }$ when we choose $n_{7}=n_{7}^{\prime
}=10,$ or an antisymmetric matter content if $n_{7}\neq n_{7}^{\prime }.$ It
is also possible to have a $G_{2}\times G_{2}$ theory with only $272$
singlet hypermultiplets.
\end{enumerate}

We have constructed theories with no perturbative anomalies, but for them to
be fully consistent theories we must go on to show that they contain no
non-perturbative anomalies. In $6D$, the global gauge anomaly \TEXTsymbol{%
\backslash}cite vanishes for all gauge groups where the 6th homotopy group
vanishes. Since $\pi _{6}\left( SO\left( 8\right) \right) =0$ the $SO\left(
8\right) $ theory is global gauge anomaly free. The $G_{2}\times %
G_{2}^{\prime }$ theories, on the other hand, have $\pi _{6}\left( G_{2}%
\times G_{2}^{\prime }\right) =Z_{3}\times Z_{3}$, but as Braden has shown 
\cite{braden}, all representations of $G_{2}$ contribute only $mod6$ to the
anomaly, and hence the global gauge anomaly vanishes in these theories too.

Finally, the global gravitational anomalies have been analyzed in $10D$ \cite
{W+A-G}, and $6D$\cite{W+A-G}\cite{BKSS}. A straightforward analysis shows
that both the $SO\left( 8\right) $ and $G_{2}\times G_{2}^{\prime }$
theories in 6D are global gravitational anomaly free.

It is obvious that the $SO\left( 8\right) $ theory can be obtained from $10D$
by compactification on $K3$ (see for example\cite{small}). The $G_{2}\times %
G_{2}^{\prime }$ theories may be more interesting. $G_{2}$ is not a simply
laced lattice, and if a corresponding string theory was to exist, its direct
construction would be non-standard. If the $G_{2}\times G_{2}^{\prime }$
theories can be obtained by compactification, then the holonomy group of $K3$
must be identified with the gauge group in a nontrivial way. One could try
the $\left( F_{4}\times G_{2}\right) \times \left( F_{4}^{\prime }\times %
G_{2}^{\prime }\right) $ irregular subgroup of $E_{8}\times E_{8}^{\prime }$%
, and attempt to extend the $SU(2)$ holonomy of $K3$ from $SU(2)$ to $F_{4}$%
, and then identify it with the diagonal $\left( F_{4}\right) _{D}=F_{4}%
\times F_{4}^{\prime }$.

\end{document}